\documentstyle[amssymb,preprint,prb,aps]{revtex}

\tightenlines

\begin{document}
\draft
\title{Optical Absorption Spectra of Bipolarons}
\author{J. T. Devreese$^{*}$, S. N. Klimin$^{**}$, and V. M. Fomin$^{**}$}
\address{Departement Natuurkunde,\\
Universiteit Antwerpen (U.I.A.), Universiteitsplein 1,\\
B-2610 Antwerpen-Wilrijk, Belgi\"e}
\date{March 09,2001}
\maketitle

\begin{abstract}
\noindent The absorption of large bipolarons is investigated using the
path-integral method. The response of a bipolaron to an external
electromagnetic field is derived in the framework of the memory-function
approach. The bipolaron optical absorption spectrum consists of a series of
relatively narrow peaks. The peculiarities of the bipolaron optical
absorption as a function of the frequency of the electromagnetic field may
be attributed to the transitions involving relaxed excited states and
scattering states of a bipolaron.
\end{abstract}

\section{Introduction}

The optical and kinetic properties of polar and ionic solids are
substantially influenced by the polaron coupling. Large polarons have been
most clearly manifested by investigations of the transport phenomena in a
magnetic field (see the recent review on polarons \cite{D96}).

When two electrons (or two holes) interact with each other simultaneously
through the Coulomb force and via the electron-phonon interaction, either
two independent polarons can occur or a bound state of two polarons --- the 
{\it bipolaron} --- can arise (see Refs. %
\onlinecite{V61,HT85,A89,V90,V91,B91} on large bipolarons and a
comprehensive review \cite{AM94} concerning small bipolarons). Whether
bipolarons are stable or not, depends on the competition between the
repulsive forces (the Coulomb interaction) and the attractive forces
(mediated through the electron-phonon interaction). Verbist {\it et al.} %
\cite{V90,V91} analyzed the large bipolaron using the Feynman path-integral
formalism \cite{F55,Feynman}. They introduced a ``phase diagram'' for the
polaron---bipolaron system on the basis of a generalization of Feynman's
trial action and showed that the Pekar-Fr\"{o}hlich coupling constant as
high as 6.8 is needed in 3D to allow for bipolaron formation. Furthermore,
in Refs. \onlinecite{V90,V91} it was shown that the large bipolaron
formation is facilitated in 2D as compared to 3D. Experimental evidences for
bipolarons, e. g. from the data on magnetization and electric conductivity
in Ti$_{4}$O$_{7}$, as well as in Na$_{0.3}$V$_{2}$O$_{5}$ and
polyacetylene, was discussed by Mott \cite{M90}.

In the framework of the renewed interest in bipolaron theory \cite%
{V61,HT85,A89,V90,B91,AM94}, a preliminary analysis of the absorption of
large bipolarons without a magnetic field was given \cite{DPVaVe94}. A
variational treatment of both spin-singlet and spin-triplet states of large
bipolarons (in 2D) and for sufficiently strong coupling in high magnetic
fields has been presented in Ref. \onlinecite{FD95}. In Ref. %
\onlinecite{BD95}, the path-integral approach for a bipolaron \cite{V90} was
generalized to the case of a bipolaron in a magnetic field for all values of
the Pekar-Fr\"{o}hlich coupling constant and for all magnetic field
strengths. It was demonstrated, \cite{BD95} that the magnetic field favors
bipolaron formation. The first investigations of the optical absorption
spectrum of large bipolarons in a magnetic field were performed in Refs. %
\onlinecite{DFB95,DF96}.

\section{Approach}

We investigate here the optical properties of a bipolaron using the
path-integral memory-function technique, developed in Refs. \onlinecite{D,4}
(see also Ref. \onlinecite{DF96}). The path-integral variational principle
gives excellent results for the free energy of the electron-phonon systems
(see, e. g., Refs. \onlinecite{Osaka,HP}). Hence, these models adequately
describe the physical properties (including the energy spectra) of the
polaron and of the bipolaron, respectively.

The Hamiltonian of two electrons, interacting with longitudinal optical (LO)
phonons and between each other in an external electric field ${\bf E}\left(
t\right) $, is 
\begin{eqnarray}
\hat{H} &=&\sum_{j=1,2}\left[ \frac{{\bf p}_{j}^{2}}{2m}+e{\bf r}_{j}\left(
t\right) \cdot {\bf E}\left( t\right) \right] +\frac{e^{2}}{\varepsilon
_{\infty }\left| {\bf r}_{1}-{\bf r}_{2}\right| }+\sum_{{\bf q}}\hbar \omega
_{{\rm LO}}\left( \hat{a}_{{\bf q}}^{+}\hat{a}_{{\bf q}}^{{}}+\frac{1}{2}%
\right)  \nonumber \\
&&+\sum_{j=1,2}\sum_{{\bf q}}\left( V_{{\bf q}}\hat{a}_{{\bf q}}^{{}}e^{{\rm %
i}{\bf q\cdot r}_{j}}+V_{{\bf q}}^{\ast }\hat{a}_{{\bf q}}^{+}e^{-{\rm i}%
{\bf q\cdot r}_{j}}\right)  \label{1}
\end{eqnarray}
with the interaction amplitudes \cite{4} 
\begin{equation}
V_{{\bf q}}=\frac{\hbar \omega _{{\rm LO}}}{{\rm i}q}\left( \frac{2\sqrt{2}%
\pi \alpha }{V}\right) ^{1/2}\left( \frac{\hbar }{m\omega _{{\rm LO}}}%
\right) ^{1/4},  \label{2}
\end{equation}
where $\alpha $\ is the electron-phonon coupling constant, $m$\ is the
electron band mass, $V$\ is the volume of the crystal, $\varepsilon _{\infty
}$\ is the high-frequency dielectric constant, $\omega _{{\rm LO}}$\ is the
LO-phonon frequency.

We consider the equation of motion for the vector function ${\bf R}\left(
t\right) $ which has the sense of the average coordinate response per one
electron, 
\begin{equation}
{\bf R}\left( t\right) \equiv \frac{1}{2}\left\langle \sum_{j=1,2}{\bf r}%
_{j}\left( t\right) \right\rangle _{S}.  \label{R}
\end{equation}
Here, the symbol of averaging denotes the path-integral average over
trajectories of the electrons (cf. Ref. \onlinecite{FHIP}) 
\begin{equation}
\left\langle F\left[ {\bf \bar{r}}\left( t\right) ,{\bf \bar{r}}^{\prime
}\left( t\right) \right] \right\rangle _{S}\equiv \frac{\int D{\bf \bar{r}}%
\left( t\right) \int D{\bf \bar{r}}^{\prime }\left( t\right) F\left[ {\bf 
\bar{r}}\left( t\right) ,{\bf \bar{r}}^{\prime }\left( t\right) \right] \exp
\left\{ \left( {\rm i}/\hbar \right) S\left[ {\bf \bar{r}}\left( t\right) ,%
{\bf \bar{r}}^{\prime }\left( t\right) \right] \right\} }{\int D{\bf \bar{r}}%
\left( t\right) \int D{\bf \bar{r}}^{\prime }\left( t\right) \exp \left\{
\left( {\rm i}/\hbar \right) S\left[ {\bf \bar{r}}\left( t\right) ,{\bf \bar{%
r}}^{\prime }\left( t\right) \right] \right\} }\quad \left[ {\bf \bar{r}%
\equiv }\left( {\bf r}_{1},{\bf r}_{2}\right) \right]  \label{Av}
\end{equation}
with the action functional 
\begin{equation}
S\left[ {\bf \bar{r}}\left( t\right) ,{\bf \bar{r}}^{\prime }\left( t\right) %
\right] =S_{e}\left[ {\bf \bar{r}}\left( t\right) \right] -S_{e}\left[ {\bf 
\bar{r}}^{\prime }\left( t\right) \right] -{\rm i}\hbar \Phi \left[ {\bf 
\bar{r}}\left( t\right) ,{\bf \bar{r}}^{\prime }\left( t\right) \right] ,
\label{Action}
\end{equation}
where $S_{e}\left[ {\bf r}_{1}\left( t\right) ,{\bf r}_{2}\left( t\right) %
\right] $\ is the action of two interacting electrons in an external
electric field, 
\begin{equation}
S_{e}\left[ {\bf \bar{r}}\left( t\right) \right] =\int_{-\infty }^{\infty
}\left\{ \sum_{j=1,2}\left[ \frac{m{\bf \dot{r}}_{j}^{2}\left( t\right) }{2}%
-e{\bf r}_{j}\left( t\right) \cdot {\bf E}\left( t\right) \right] -\frac{%
e^{2}}{\varepsilon _{\infty }\left| {\bf r}_{1}\left( t\right) -{\bf r}%
_{2}\left( t\right) \right| }\right\} dt,  \label{Se}
\end{equation}
and $\Phi \left[ {\bf \bar{r}}\left( t\right) ,{\bf \bar{r}}^{\prime }\left(
t\right) \right] $ is the ``influence phase'' of the phonon subsystem \cite%
{DF96} 
\begin{eqnarray}
\Phi \left[ {\bf \bar{r}}\left( t\right) ,{\bf \bar{r}}^{\prime }\left(
t\right) \right] &=&-\sum_{{\bf q}}\frac{\left| V_{{\bf q}}\right| ^{2}}{%
\hbar ^{2}}\int\limits_{-\infty }^{\infty }dt\int\limits_{-\infty
}^{t}dt^{\prime }\left[ \rho _{{\bf q}}\left( t\right) -\rho _{{\bf q}%
}^{\prime }\left( t\right) \right]  \nonumber \\
&&\times \left[ T_{\omega _{{\bf q}}}^{\ast }\left( t-t^{\prime }\right)
\rho _{-{\bf q}}\left( t^{\prime }\right) -T_{\omega _{{\bf q}}}\left(
t-t^{\prime }\right) \rho _{-{\bf q}}^{\prime }\left( t^{\prime }\right) %
\right] .
\end{eqnarray}
This influence phase describes both a retarded interaction between different
electrons and a retarded self-interaction of each electron due to the
elimination of the phonon coordinates. It contains the Fourier components of
the electron density operator 
\begin{equation}
\rho _{{\bf q}}=\sum_{j=1,2}{\rm e}^{{\rm i}{\bf qr}_{j}}
\end{equation}
and the phonon Green's function 
\begin{equation}
T_{\omega }\left( t\right) =\frac{e^{{\rm i}\omega t}}{1-e^{-\beta \omega }}+%
\frac{e^{-{\rm i}\omega t}}{e^{\beta \omega }-1},\quad \beta \equiv \frac{%
\hbar }{k_{B}T},
\end{equation}
$T$ is the temperature.

We have calculated the optical absorption coefficient for a bipolaron in the
memory-function approach \cite{DF96,D,4}, 
\begin{equation}
\Gamma \left( \omega \right) =-\frac{4\pi }{cn}\frac{n_{0}e^{2}}{m}\frac{%
\omega {\mbox {Im}}T\left( \omega \right) }{\left[ \omega ^{2}-{\mbox {Re}}%
T\left( \omega \right) \right] ^{2}+\left[ {\mbox {Im}}T\left( \omega
\right) \right] ^{2}},  \label{Kw}
\end{equation}
where $c$\ is the velocity of light, $n$\ is the refractive index of the
crystal, $n_{0}$\ is the electron density. The memory function $T\left(
\omega \right) $ has the form \cite{DF96}: 
\begin{equation}
T\left( \omega \right) =\sum_{{\bf q}}\frac{\left| V_{{\bf q}}\right|
^{2}q^{2}}{3\hbar }\int\limits_{0}^{\infty }\left( {\rm e}^{{\rm i}\omega
t}-1\right) \mbox{Im}\left[ T_{\omega _{{\rm LO}}}^{\ast }\left( t\right)
\left\langle \rho _{{\bf q}}\left( t\right) \rho _{-{\bf q}}\left( 0\right)
\right\rangle _{S_{0}}\right] \,dt.  \label{T}
\end{equation}
It is expressed in terms of the two-point correlation function $\left\langle
\rho _{{\bf q}}\left( t\right) \rho _{-{\bf q}}\left( 0\right) \right\rangle
_{S_{0}}$ of the electron density operators. The correlation function is
calculated as the path-integral average [cf. Eq. (\ref{Av})] with the model
action functional $S_{0}\left[ {\bf \bar{r}}\left( t\right) ,{\bf \bar{r}}%
_{f}\left( t\right) ,{\bf \bar{r}}^{\prime }\left( t\right) ,{\bf \bar{r}}%
_{f}^{\prime }\left( t\right) \right] ${\bf : } 
\begin{equation}
\left\langle F\right\rangle _{S_{0}}\equiv \frac{\int D{\bf \bar{r}}\left(
t\right) \int D{\bf \bar{r}}^{\prime }\left( t\right) \int D{\bf \bar{r}}%
_{f}\left( t\right) \int D{\bf \bar{r}}_{f}^{\prime }\left( t\right) F\exp
\left\{ \left( {\rm i}/\hbar \right) S_{0}\left[ {\bf \bar{r}}\left(
t\right) ,{\bf \bar{r}}_{f}\left( t\right) ,{\bf \bar{r}}^{\prime }\left(
t\right) ,{\bf \bar{r}}_{f}^{\prime }\left( t\right) \right] \right\} }{\int
D{\bf \bar{r}}\left( t\right) \int D{\bf \bar{r}}^{\prime }\left( t\right)
\int D{\bf \bar{r}}_{f}\left( t\right) \int D{\bf \bar{r}}_{f}^{\prime
}\left( t\right) \exp \left\{ \left( {\rm i}/\hbar \right) S_{0}\left[ {\bf 
\bar{r}}\left( t\right) ,{\bf \bar{r}}_{f}\left( t\right) ,{\bf \bar{r}}%
^{\prime }\left( t\right) ,{\bf \bar{r}}_{f}^{\prime }\left( t\right) \right]
\right\} }.  \label{AvS0}
\end{equation}
The model system consists of two electrons harmonically interacting with two
fictitious particles and between each other. The model action functional $%
S_{0}\left[ {\bf \bar{r}}\left( t\right) ,{\bf \bar{r}}_{f}\left( t\right) ,%
{\bf \bar{r}}^{\prime }\left( t\right) ,{\bf \bar{r}}_{f}^{\prime }\left(
t\right) \right] $\ has the form 
\begin{equation}
S_{0}\left[ {\bf \bar{r}}\left( t\right) ,{\bf \bar{r}}_{f}\left( t\right) ,%
{\bf \bar{r}}^{\prime }\left( t\right) ,{\bf \bar{r}}_{f}^{\prime }\left(
t\right) \right] =\int_{-\infty }^{\infty }\left[ L_{0}\left( {\bf \dot{\bar{%
r}},\dot{\bar{r}}}_{f}{\bf ,\bar{r},\bar{r}}_{f}\right) -L_{0}\left( {\bf 
\dot{\bar{r}}}^{\prime }{\bf ,\dot{\bar{r}}}_{f}^{\prime }{\bf ,\bar{r}}%
^{\prime }{\bf ,\bar{r}}_{f}^{\prime }\right) \right] dt,  \label{S0}
\end{equation}
with the Lagrangian 
\begin{eqnarray}
L_{0}\left( {\bf \dot{\bar{r}},\dot{\bar{r}}}_{f}{\bf ,\bar{r},\bar{r}}%
_{f}\right) &=&\sum_{j=1,2}\left( \frac{m{\bf \dot{r}}_{j}^{2}\left(
t\right) }{2}+\frac{M{\bf \dot{r}}_{fj}^{2}\left( t\right) }{2}\right) -%
\frac{k}{2}\sum_{j=1,2}\left( {\bf r}_{j}-{\bf r}_{fj}\right) ^{2}  \nonumber
\\
&&-\frac{k^{\prime }}{2}\left[ \left( {\bf r}_{1}-{\bf r}_{f2}\right)
^{2}+\left( {\bf r}_{2}-{\bf r}_{f1}\right) ^{2}\right] +\frac{K}{2}\left( 
{\bf r}_{1}-{\bf r}_{2}\right) ^{2},  \label{L0}
\end{eqnarray}
where $M$\ is the mass of a fictitious particle, $k,$\ $k^{\prime }$\ and $K$%
\ are the elastic constants. The oscillator potentials in the Lagrangian (%
\ref{L0}) imitate the electron-phonon interaction and the electron-electron
Coulomb repulsion. In Refs. \onlinecite{V90,V91}, the aforesaid model has
been first introduced in order to calculate the bipolaron ground-state
energy by the Jensen-Feynman variational method \cite{F55,Feynman}. The
variational functional of Refs. \onlinecite{V90,V91} for the bipolaron free
energy contains four variational parameters: $M,$\ $k,$\ $k^{\prime }$\ and $%
K$. It is convenient to use instead of them four other independent
parameters: (i) three eigenfrequencies of the internal bipolaron motion $%
\Omega _{i}$\ $\left( i=1,2,3\right) $\ (see Ref. \onlinecite{V91}), 
\begin{eqnarray}
\Omega _{1} &=&\frac{m+M}{mM}\left( k+k^{\prime }\right) ,  \label{par1} \\
\Omega _{2,3}^{2} &=&\frac{1}{2}\left\{ \frac{m+M}{mM}\left( k+k^{\prime
}\right) -\frac{2K}{m}\pm \left[ \left( \frac{M-m}{mM}\left( k+k^{\prime
}\right) -\frac{2K}{m}\right) ^{2}+\frac{4}{mM}\left( k-k^{\prime }\right)
^{2}\right] ^{1/2}\right\} ,  \label{par2}
\end{eqnarray}
and (ii) the frequency 
\begin{equation}
v=\left( \frac{k+k^{\prime }}{M}\right) ^{1/2},  \label{par3}
\end{equation}
which is analogous to the Feynman parameter $w$\ in the single-polaron
problem \cite{Feynman}. It is seen from Eqs. (\ref{par1}) to (\ref{par3}),
that the inequality $\Omega _{1}\geq \Omega _{2}\geq v\geq \Omega _{3}$\ is
fulfilled for the variational frequency parameters. By the variational
procedure of Ref. \onlinecite{V91}, the optimal values of those variational
parameters are found for the physical system of two electrons interacting
with the phonon field and between each other.

In the zero-temperature limit $T=0,$ we have derived from Eq. (\ref{T}) the
analytical expressions for the real and imaginary parts of the memory
function for a bipolaron in three dimensions: 
\begin{eqnarray}
&&{\mbox {Re}}T\left( \omega \right) =\frac{4\sqrt{2}\alpha }{3}%
\sum_{n=0}^{\infty }\sum_{k=0}^{\infty }\sum_{l=0}^{\infty }\frac{\left[
1+\left( -1\right) ^{k+l}\right] }{n!k!l!}s_{1}^{n}s_{2}^{k}s_{3}^{l} 
\nonumber \\
&\times &\left[ f_{n+k+l}\left( \omega -1-n\Omega _{1}-k\Omega _{2}-l\Omega
_{3}\right) +f_{n+k+l}\left( -\omega -1-n\Omega _{1}-k\Omega _{2}-l\Omega
_{3}\right) \right.  \nonumber \\
&-&2\left. f_{n+k+l}\left( -1-n\Omega _{1}-k\Omega _{2}-l\Omega _{3}\right) 
\right] ,  \label{ReT}
\end{eqnarray}
\begin{eqnarray}
&&\mbox{Im}T\left( \omega \right) =-\frac{4\sqrt{2}\alpha }{3}%
\sum_{n=0}^{\infty }\sum_{k=0}^{\infty }\sum_{l=0}^{\infty }\frac{\left[
1+\left( -1\right) ^{k+l}\right] }{n!k!l!}s_{1}^{n}s_{2}^{k}s_{3}^{l} 
\nonumber \\
&\times &\left[ g_{n+k+l}\left( \omega -1-n\Omega _{1}-k\Omega _{2}-l\Omega
_{3}\right) -g_{n+k+l}\left( -\omega -1-n\Omega _{1}-k\Omega _{2}-l\Omega
_{3}\right) \right] .  \label{ImT}
\end{eqnarray}
From here on, we use the system of units where $\hbar =1,$\ $m=1,$\ $\omega
_{{\rm LO}}=1.$\ The factors $s_{i}$\ ($i=1,2,3$) in Eqs. (\ref{ReT}), (\ref%
{ImT}) have the sense of oscillator strengths corresponding to the
eigenmodes of the internal motion of a bipolaron. The aforesaid factors are
expressed through variational frequency parameters of the bipolaron model
functional (\ref{par1}) to (\ref{par3}), 
\begin{equation}
s_{1}=\frac{\Omega _{1}^{2}-\nu ^{2}}{\Omega _{1}^{3}},\quad s_{2}=\frac{%
\Omega _{2}^{2}-\nu ^{2}}{\Omega _{2}\left( \Omega _{2}^{2}-\Omega
_{3}^{2}\right) },\quad s_{3}=\frac{\nu ^{2}-\Omega _{3}^{2}}{\Omega
_{3}\left( \Omega _{2}^{2}-\Omega _{3}^{2}\right) }.
\end{equation}
The functions $f_{n}\left( \omega \right) $ and $g_{n}\left( \omega \right) $
are given by the expressions: 
\begin{equation}
f_{n}\left( \omega \right) =\left( -1\right) ^{n}\frac{\left| \omega \right|
^{n+\frac{1}{2}}}{B^{n+\frac{3}{2}}}e^{-\frac{A\omega }{B}}\Theta \left(
-\omega \right) -\frac{\Gamma \left( n+\frac{1}{2}\right) }{\pi A^{n+\frac{1%
}{2}}B}\left. _{1}F_{1}\left( 1,\frac{1}{2}-n,-\frac{A\omega }{B}\right)
\right. ,  \label{real}
\end{equation}
\begin{equation}
g_{n}\left( \omega \right) =\frac{\omega ^{n+\frac{1}{2}}}{B^{n+\frac{3}{2}}}%
e^{-\frac{A\omega }{B}}\Theta \left( \omega \right) ,  \label{im}
\end{equation}
with the parameters 
\begin{equation}
A=\frac{\mu }{m\Omega _{1}}+\frac{a^{2}}{\Omega _{2}}+\frac{b^{2}}{\Omega
_{3}},\quad \mu =\frac{\Omega _{1}^{2}-\nu ^{2}}{\Omega _{1}^{2}},\quad a=%
\sqrt{\frac{\Omega _{2}^{2}-\nu ^{2}}{\Omega _{2}^{2}-\Omega _{3}^{2}}}%
,\quad b=\sqrt{\frac{\nu ^{2}-\Omega _{3}^{2}}{\Omega _{2}^{2}-\Omega
_{3}^{2}}},\quad B=\frac{\nu ^{2}}{\Omega _{1}^{2}}.  \label{par}
\end{equation}

The optical conductivity for a 2D bipolaron is related to that for a 3D
bipolaron, which is given by the formulas (\ref{ReT}) and (\ref{ImT}), by
the scaling relation (cf. Ref. \onlinecite{D96}): 
\begin{equation}
\mathop{\rm Re}%
\sigma _{2{\rm D}}(\omega ,\alpha )=%
\mathop{\rm Re}%
\sigma _{3{\rm D}}\left( \omega ,\frac{3\pi }{4}\alpha \right) .
\label{scaling}
\end{equation}

It follows from Eqs. (\ref{ReT}) and (\ref{ImT}) that the eigenfrequencies $%
\Omega _{2}$\ and $\Omega _{3}$\ appear in the optical absorption spectra
only in such combinations $\left( k\Omega _{2}+l\Omega _{3}\right) $\ that $%
\left( k+l\right) $\ is an even integer. This selection rule is determined
by the symmetry of these eigenmodes (a schematic picture of the internal
motion of the bipolaron model system see, e. g., in Ref. \onlinecite{DF96}).
The only normal coordinates of eigenmodes with the frequencies $\Omega _{2}$%
\ and $\Omega _{3}$\ (let us denote these coordinates as vectors $Q_{2}$\
and $Q_{3}$) are antisymmetric with respect to the permutation of electrons $%
r_{1}\rightleftarrows r_{2}.$\ Both the exact and model Lagrangians are
symmetric with respect to this permutation. As a result of this symmetry, we
obtain the selection rule 
\begin{equation}
\left\langle \prod_{i=1,2,\dots }\prod_{k=x,y,z}\left[ Q_{2k}^{m_{ki}}\left(
t_{i}\right) Q_{3k}^{m_{ki}^{\prime }}\left( t_{i}^{\prime }\right) \right]
\right\rangle _{S_{0}}=0,\quad {\rm when}\;\sum_{i}\sum_{k=x,y,z}\left(
m_{ki}+m_{ki}^{\prime }\right) ={\rm odd}.  \label{Selection}
\end{equation}
Hence, only the combinations $\left( \prod_{i=1,2,\dots }\prod_{k=x,y,z}%
\left[ Q_{2k}^{m_{ki}}\left( t_{i}\right) Q_{3k}^{m_{ki}^{\prime }}\left(
t_{i}^{\prime }\right) \right] \right) $\ with an even number $%
n=\sum_{i}\sum_{k=x,y,z}\left( m_{ki}+m_{ki}^{\prime }\right) $\ can give a
nonzero contribution into the average $\left\langle \rho _{{\bf q}}\left(
t\right) \rho _{-{\bf q}}\left( 0\right) \right\rangle _{S_{0}}$\ of Eq. (%
\ref{T}). The operator of an oscillator coordinate $Q_{jk}$\ describes
quantum transitions between energy levels of the corresponding oscillator
with the frequency $\Omega _{j}$\ $\left( j=1,2,3\right) $. Consequently,
the aforesaid combinations of coordinates provide the transitions, which
change the energy of the model system by values $\left( k\Omega _{2}+l\Omega
_{3}\right) $\ with even $\left( k+l\right) .$

\section{Optical absorption spectra}

In order to give a physical interpretation of the calculated bipolaron
optical absorption spectra, we refer first to a description of those for a
polaron in Refs. \onlinecite{D,4}. In the polaron optical absorption
spectra, for intermediate and large $\alpha ,$\ there is an intense
(zero-phonon) absorption peak corresponding to a transition from the ground
state to the first relaxed excited state (RES). The relaxed excited state is
created if the electron in the polaron is excited while the lattice readapts
to a new electronic configuration \cite{4}. A shoulder at the low-frequency
side of the RES peak is attributed to one-phonon transitions through the
scattering states (ScS): excitations of the polaron system characterized by
the presence of a finite number of real phonons along with the polaron \cite%
{4}. A broad peak, positioned at a higher frequency than the RES peak, is
attributed to transitions to Franck-Condon (FC) states: internal excited
polaron states for which the lattice polarization is that of the polaron in
its ground state \cite{4}. This picture is well-founded physically and is in
agreement with the prediction of the aforesaid peaks, which was formulated
within the strong-coupling approach \cite{KED}.

In Fig. 1, the real $\mbox{Re}T\left( \omega \right) $\ and imaginary $%
\mbox{Im}T\left( \omega \right) $\ parts of the memory function for a 3D
(2D) bipolaron are plotted as a function of frequency $\omega $\ for $\alpha
=7$\ ($2.971$). These values of $\alpha $\ are close to the minimal possible 
$\alpha $\ for the bipolaron formation: $\alpha _{\min }=6.8$\ ($2.9$)(see
Ref. \onlinecite{DF96}). The peaks of the imaginary part of the memory
function are positioned near the points (explicitly indicated in Fig. 1) 
\begin{equation}
\omega _{nkl}\equiv n\Omega _{1}+k\Omega _{2}+l\Omega _{3}+1,
\label{position}
\end{equation}
where $n,$\ $k,$\ $l$\ are the non-negative integers ($k+l=$\ even integer).

Analogously to the case of polaron optical absorption \cite{4}, the peaks of
the bipolaron optical absorption spectra (Fig. 2) correspond to: (i) peaks
of $\mbox{Im}T\left( \omega \right) $\ and (ii) roots of the equation $%
\omega ^{2}-%
\mathop{\rm Re}%
T\left( \omega \right) =0$\ (the crossing points of solid and dotted curves
in the panel ``a'' of Fig. 1). Following the physical interpretation
developed in Refs. \onlinecite{D,4}, we suggest that the peaks corresponding
to zeros of $\left[ \omega ^{2}-%
\mathop{\rm Re}%
T\left( \omega \right) \right] $\ can be attributed to transitions to {\em %
bipolaron RES.} Due to a larger number of the internal degrees of freedom
for a bipolaron, than that for a polaron, there are several types of RES for
a bipolaron.

The positions of peaks of the optical absorption coefficient, corresponding
to those of $\mbox{Im}T\left( \omega \right) ,$\ are close to the values $%
\omega _{nkl}$\ determined by Eq. (\ref{position}), so that $\omega _{nkl}$\
have the sense of the frequencies of transitions from the ground state to
certain internal states of a bipolaron. Following the physical
interpretation developed in Refs. \onlinecite{D,4}, we can attribute these
peaks to transitions into\ {\em Franck-Condon} (FC) bipolaron states (in the
case when at least one of the numbers $\left( n,k,l\right) \neq 0,$\ i. e.,
for {\em excited} states of a bipolaron). Within the same picture, we
suggest that the low-intensity peak at $\omega =1$\ is provided by the
transition into the lowest {\em scattering} state of a bipolaron. It is
worth mentioning, that the main resonances in oscillator strength are RES.

In order to classify peaks of the bipolaron optical absorption spectra, we
have calculated transition frequencies to several bipolaron RES and FC
states within the strong-coupling (adiabatic) approach. The results are
shown in Table 1. The transitions are classified with respect to the
symmetry of states of the internal bipolaron motion and of the motion of a
bipolaron as a whole (labelled, respectively, by small and capital letters
within standard denotations of states with different orbital moments). ``$S$%
'' and ``$s$'' denote the states with the orbital momentum $l=0,$ while ``$P$%
'' and ``$p$'' are those with $l=1.$ The trial wave functions are of the
same type as those used for the treatment of polarons in Ref. \onlinecite{4}.

\bigskip

\begin{center}
{\bf Table 1. Transition frequencies (in units of the LO phonon frequency)
to bipolaron RES and FC states calculated within the adiabatic
strong-coupling theory.}

\medskip

\begin{tabular}{|c|ccc|ccc|}
\hline
$\alpha $ &  & RES &  &  & FC &  \\ \cline{2-7}
& $pS$ & \multicolumn{1}{|c}{$sP$} & \multicolumn{1}{|c|}{$pP$} & $pS$ & 
\multicolumn{1}{|c}{$sP$} & \multicolumn{1}{|c|}{$pP$} \\ \hline
$7$ & $2.2171$ & \multicolumn{1}{|c}{$5.9877$} & \multicolumn{1}{|c|}{$%
5.7282 $} & $5.4633$ & \multicolumn{1}{|c}{$9.9396$} & \multicolumn{1}{|c|}{$%
12.5210 $} \\ 
$8$ & $2.8957$ & \multicolumn{1}{|c}{$7.8207$} & \multicolumn{1}{|c|}{$%
7.4818 $} & $7.1328$ & \multicolumn{1}{|c}{$13.0008$} & \multicolumn{1}{|c|}{%
$16.3627$} \\ 
$9$ & $3.6648$ & \multicolumn{1}{|c}{$9.8981$} & \multicolumn{1}{|c|}{$%
9.4686 $} & $8.9364$ & \multicolumn{1}{|c}{$16.2629$} & \multicolumn{1}{|c|}{%
$20.4673$} \\ 
$10$ & $4.5243$ & \multicolumn{1}{|c}{$12.220$} & \multicolumn{1}{|c|}{$%
11.6895$} & $11.1019$ & \multicolumn{1}{|c}{$20.1550$} & 
\multicolumn{1}{|c|}{$25.4065$} \\ \hline
\end{tabular}
\end{center}

\bigskip

\noindent Comparing the peak positions of Fig. 2 with the results of Table 1
(which are obtained in the framework of the adiabatic approach, and
therefore can be applied only approximately), we have attributed several
optical absorption peaks to transitions to RES and FC states of a definite
symmetry, as shown at the figure. It should be mentioned, that every RES
peak has a lower frequency, than the corresponding FC peak, since the
lattice relaxation leads to a lowering of the energy of the electron-phonon
system.

The linewidth of peaks of bipolaron optical absorption spectra is determined
by the bipolaron recoil in the scattering process. Every FC peak is
described by a function $g_{n+k+l}\left( \omega -\omega _{nkl}\right) $\ of
Eq. (\ref{ImT}), where $g_{n}\left( \omega \right) $\ is given by Eq. (\ref%
{im}). Consequently, the linewidth of a FC peak can be estimated, using the
following characteristic of the function (\ref{im}): 
\begin{equation}
\sigma \left[ g_{n}\left( \omega \right) \right] =\sqrt{\left\langle \omega
^{2}\right\rangle -\left\langle \omega \right\rangle ^{2}},\quad
\left\langle \omega ^{k}\right\rangle \equiv \frac{\int_{0}^{\infty }\omega
^{k}g_{n}\left( \omega \right) d\omega }{\int_{0}^{\infty }g_{n}\left(
\omega \right) d\omega }.  \label{bbbb}
\end{equation}
Performing the integrations over $\omega $\ analytically, we find that 
\[
\sigma \left[ g_{n}\left( \omega \right) \right] =\frac{B}{A}\left( n+\frac{3%
}{2}\right) ^{1/2}, 
\]
where the parameters $A$\ and $B$\ are given by Eq. (\ref{par}). The ratio $%
\frac{A}{B}$\ is of the same order as $\left[ 2\left( M+1\right) \right] ,$\
which is used for estimation of the bipolaron effective mass \cite{Bipwire}.
Therefore, the characteristic linewidth of a FC peak corresponding to a
definite frequency $\omega =\omega _{nkl}$\ can be qualitatively estimated
as 
\[
\Gamma _{nkl}^{\left( FC\right) }\propto \frac{1}{2\left( M+1\right) }\left(
n+k+l+3/2\right) ^{1/2}. 
\]
Since the bipolaron effective mass is larger than that of a polaron, the
bipolaron optical absorption spectrum consists of a series of peaks which
are narrower in comparison with those of the polaron absorption spectrum.

Let us define $N\left( \omega ,\Delta \omega \right) $\ as the number of
possible combinations of $\left( n,k,l\right) $\ for which the frequencies $%
\left\{ \omega _{nkl}\right\} $\ occupy the interval $\omega \leqslant
\omega _{nkl}\leqslant \omega +\Delta \omega .$\ The higher is $\omega $\
(at a fixed interval length $\Delta \omega $), the larger is $N\left( \omega
,\Delta \omega \right) .$\ Hence, the density of the FC peaks increases with
increasing frequency. Starting from a definite frequency range (where $%
N\left( \omega ,\Delta \omega \right) \propto \Gamma _{FC}\left( \omega
\right) \Delta \omega $), FC peaks merge into a continuous band.

In the limiting case $\omega \gg 1,$\ the imaginary part of the memory
function (\ref{ImT}) asymptotically behaves as 
\begin{equation}
\left. 
\mathop{\rm Im}%
T\left( \omega \right) \right| _{\omega \gg 1}\propto -\frac{2\alpha }{3}%
\omega ^{1/2}.
\end{equation}
Hence, the bipolaron absorption coefficient in the limiting case of high
frequencies has the same asymptotic behavior as the polaron absorption
coefficient \cite{D,4} (this asymptotic behavior was also derived in Ref. %
\onlinecite{Gur} for the weak-coupling polaron optical absorption): 
\begin{equation}
\left. \Gamma \left( \omega \right) \right| _{\omega \gg 1}\propto \omega
^{-5/2}.  \label{ppp}
\end{equation}
This behavior has a clear physical explanation: in the high-frequency limit,
only one-phonon scattering processes give a contribution into the optical
absorption by the electron-phonon system. In this limit, the optical
absorption coefficients for bipolarons and polarons at arbitrary coupling
strength are described by {\em one and the same} asymptotic dependence. It
was emphasized in Refs. \onlinecite{D,4,Gur}, that the frequency dependence (%
\ref{ppp}) of the polaron optical absorption coefficient at $\omega \gg 1$\
differs from the Drude-like dependence $\left. \Gamma ^{\left( {\rm Drude}%
\right) }\left( \omega \right) \right| _{\omega \gg 1}\propto \omega ^{-2}$.

For the bipolaron, the optical absorption spectrum turns into a continuous
band at substantially higher frequencies, than for the polaron, due to a
comparatively large effective mass of a bipolaron. This frequency region is
not presented in Fig. 2. The bipolaron absorption coefficient in this region
is very small when compared with that shown in Fig. 2.

\section{Sum rules and conclusions}

From the general analytical properties of the memory function, for the real
part of the optical conductivity per one electron 
\begin{equation}
\mbox{Re}\sigma \left( \omega \right) =-\frac{e^{2}}{m}\frac{\omega \mbox{Im}%
T\left( \omega \right) }{\left[ \omega ^{2}-\mbox{Re}T\left( \omega \right) %
\right] ^{2}+\left[ \mbox{Im}T\left( \omega \right) \right] ^{2}}
\end{equation}
the following sum rule was derived in Ref. \onlinecite{LSD}: 
\begin{equation}
\frac{\pi e^{2}}{m^{\ast }}+\int_{\varepsilon \rightarrow +0}^{\infty }%
\mbox{Re}\sigma \left( \omega \right) d\omega =\frac{\pi e^{2}}{2m},
\label{srule}
\end{equation}
where $m^{\ast }$\ is the polaron effective mass. For the bipolaron optical
absorption, the same equation (\ref{srule}) is valid with $m^{\ast }$\ the
bipolaron effective mass. Our numerical test of the sum rule (\ref{srule})
has confirmed, that Eq. (\ref{srule}) is fulfilled within the chosen
relative accuracy (up to 10$^{-3}$){\bf .}

In Ref. \onlinecite{2}, the ground state theorem for a {\em polaron} has
been derived which relates the polaron ground state energy to the first
moment of the optical absorption spectra. The extension of this theorem to
the bipolaron can be performed in the same way as in Ref. \onlinecite{2}.
The analog of Eq. (25) from Ref. \onlinecite{2} for a {\em bipolaron} is 
\begin{equation}
E_{0}\left( \alpha _{2},\eta \right) -E_{0}\left( \alpha _{1},\eta \right) =-%
\frac{3m\hbar }{\pi e^{2}}\int_{\alpha _{1}}^{\alpha _{2}}\frac{d\alpha }{%
\alpha }\int_{0}^{\infty }\mbox{Im}\chi _{jj}\left( \omega ,\alpha ,\eta
\right) d\omega ,  \label{rule1}
\end{equation}
where $\alpha _{1}$ and $\alpha _{2}$ are two arbitrary values of $\alpha ,$ 
$\eta =\varepsilon _{\infty }/\varepsilon _{0}$ is the ratio of the
high-frequency and static dielectric constants, $\chi _{jj}\left( \omega
,\alpha ,\eta \right) $ is the current-current correlation function \cite{2}%
. For a bipolaron, we choose $\eta $ and both $\alpha _{1}$ and $\alpha _{2}$
within the region of the bipolaron stability. The function $\mbox{Im}\chi
_{jj}\left( \omega \right) $ for a bipolaron is expressed through the memory
function $T\left( \omega \right) $ as 
\begin{equation}
\mbox{Im}\chi _{jj}\left( \omega \right) =\frac{2\omega ^{2}\mbox{Im}T\left(
\omega \right) }{\omega ^{4}-2\omega ^{2}\mbox{Re}T\left( \omega \right)
+\left| T\left( \omega \right) \right| ^{2}}.
\end{equation}
In Fig. 3, the ground-state theorem (\ref{rule1}) is illustrated
numerically. The solid curve shows the difference between the ground state
energies $E_{0}\left( \alpha _{2},\eta \right) -E_{0}\left( \alpha _{1},\eta
\right) $ (calculated by the variational method using the variational
functional of Ref. \onlinecite{V91}) for two values of $\alpha $ at $\eta
=0. $ The value $\alpha _{1}$ is taken $\alpha _{1}=7.$ The dashed curve
shows the values of the right-hand side of Eq. (\ref{rule1}). As seen from
Fig. 3, there exists an excellent agreement between the quantities entering
the independently calculated left-hand and the right-hand sides of the
equation (\ref{rule1}).

In the present paper, we have treated the bipolaron optical absorption
within the memory-function path-integral approach. The derived optical
absorption spectra demonstrate a rich structure of relatively narrow peaks,
which is clearly related to the internal states of a bipolaron. We have
attributed those peaks of the bipolaron optical absorption spectra to
transitions from the ground state to (i) scattering states, (ii) relaxed
excited states, (iii) Franck-Condon states. Every RES peak is shifted to a
lower frequency with respect to the corresponding FC peak. This
interpretation is performed within a unified approach to the polaron and
bipolaron problems, which takes its origin in Refs. \onlinecite{D,4}. The
sum rules, developed for the polaron optical conductivity \cite{LSD,2}, are
extended to the case of a bipolaron.

\acknowledgements This work has been supported by the BOF NOI (UA-UIA), GOA
BOF UA 2000, IUAP, FWO-V. projects G.0287.95, 9.0193.97, and the W.O.G.
WO.025.99N (Belgium).


\newpage

\begin{center}
{\bf Figure captions}
\end{center}

\bigskip

Fig. 1. Real [panel ``a''] and imaginary [panel ``b''] parts of the memory
function $T\left( \omega \right) $ for the optical absorption coefficient of
a 3D (2D) bipolaron for $\alpha =7$ (2.971), $\eta =0.0037$. The optimal
values of the variational frequencies are indicated in the figure.

\bigskip

Fig. 2. Optical absorption spectra of a 3D (2D) bipolaron for $\alpha =7$
(2.971), $\eta =0.0037$ [panel ``a''], for $\alpha =8$ (3.395), $\eta =0.023$
[panel ``b''], and for $\alpha =9$\ (3.820), $\eta =0.023$\ [panel ``c''].

\bigskip

Fig. 3. Illustration of the ground-state theorem (\ref{rule1}) for the
bipolaron optical absorption. The index ``1'' denotes the ground state
energies calculated by the variational method [the left-hand side of Eq. (%
\ref{rule1})], while the index ``2'' denotes the ground state energies
calculated by integration of the bipolaron optical absorption spectra [the
right-hand side of Eq. (\ref{rule1})].

\end{document}